\documentclass[useAMS,usenatbib,letterpaper]{mn2e}
\usepackage{amsmath,amssymb,graphicx,psfig,times}

\def\deg{^{\circ}}
\def\arcmin{^{\prime}}
\def\arcsec{^{\prime\prime}}
\def\hmpc{h^{-1}Mpc}
\def\kps{kms^{-1}}

\title[Photometric Selection of Emission Line Galaxies]{Photometric Selection of Emission Line Galaxies, Clustering Analysis and a Search for the ISW effect.}
\author[R. Bielby et al] {Rich Bielby$^{1,2}$\thanks{E-mail:bielby@iap.fr (RMB)}, T. Shanks$^{1}$, U. Sawangwit$^{1}$, S. M. Croom$^3$, Nicholas P. Ross$^{1,4}$
\newauthor and D. A. Wake$^{1}$ \\
{\small $^{1}$Dept of Physics, Durham University, South Road, Durham, DH1 3LE, UK}\\
{\small $^{2}$Institut dÕAstrophysique de Paris, UMR 7095 CNRS, UniversitŽ Pierre et Marie Curie, 98bis boulevard Arago, 75014 Paris, France}\\
{\small $^{3}$Sydney Institute for Astronomy, School of Physics, University of Sydney, NSW 2006, Australia}\\
{\small $^{4}$Dept of Astronomy and Astrophysics, Pennsylvania State University, 525 Davey Laboratory, University Park, PA 16802, USA}}

\begin{document}

\pagerange{\pageref{firstpage}--\pageref{lastpage}} \pubyear{2009}

\maketitle

\label{firstpage}

\begin{abstract}
We investigate the use of simple colour cuts applied to the Sloan Digital Sky Survey (SDSS) optical imaging to perform photometric selections of emission line galaxies out to $z<1$. Our selection is aimed at discerning three separate redshift ranges: $0.2\lesssim z\lesssim0.4$, $0.4\lesssim z\lesssim0.6$ and $0.6\lesssim z\lesssim1.0$. We calibrate the selections using data taken by the COMBO-17 survey in a single field (S11), which is covered by the SDSS imaging. We thus perform colour-cuts using the SDSS g, r and i bands and obtain mean photometric redshifts of $\overset{\_}{z}_{low}=0.32\pm0.08$, $\overset{\_}{z}_{mid}=0.44\pm0.12$ and $\overset{\_}{z}_{hi}=0.65\pm0.21$. We further calibrate our high redshift selection using spectroscopic observations with the AAOmega spectrograph on the 4m Anglo-Australian Telescope (AAT), observing $\approx50-200$ galaxy candidates in 4 separate fields. With just 1-hour of integration time and with seeing of $\approx1.6\arcsec$, we successfully determined redshifts for $\approx65\%$ of the targeted candidates. We compare our spectroscopic redshifts to the photometric redshifts from the COMBO-17 survey and find reasonable agreement between the two. We calculate the angular correlation functions of these samples and find correlation lengths of $r_0=2.64_{-0.08}^{+0.05} h^{-1}Mpc$, $r_0=3.62\pm0.06 h^{-1}Mpc$ and $r_0=5.88\pm 0.12h^{-1}Mpc$ for the low, mid and high redshift samples respectively. Comparing these results with predicted dark matter clustering, we estimate the bias parameter for each sample to be $b=0.70$, $b=0.92$ and $b=1.46$. We calculate the 2-point redshift-space correlation function at $z\approx0.6$ and find a clustering amplitude of $s_o=6.4\pm0.8 h^{-1}Mpc$. Finally, we use our photometric sample to search for the Integrated Sachs-Wolfe signal in the WMAP 5yr data. We cross-correlate our three redshift samples with the WMAP W, V, Q and K bands and find an overall trend for a positive signal similar to that expected from models. However, the signal in each is relatively weak with the results in the WMAP W-band of $w_{Tg}(<100\arcmin)=0.25\pm0.27 \mu K$, $0.17\pm0.20 \mu K$ and $0.17\pm0.16 \mu K$ for the low, mid and high redshift samples respectively. Combining all three galaxy samples we find a signal of $w_{Tg}(<100\arcmin)=0.20\pm0.12 \mu K$ in the WMAP W-band, a significance of $1.7\sigma$. However, in testing for systematics where the WMAP data is rotated with respect to the ELG sample, we found similar results at several different rotation angles, implying the apparent signal may be produced by systematic effects.
\end{abstract}


\section{Introduction}

Imaging surveys are currently in the process of mapping out a vast region of the Universe over a large range of the electromagnetic spectrum. The pace-setter in recent years is the Sloan Digital Sky Survey (SDSS, \citealt{york00}), which now provides (as of DR6, \citealt{adelmanmccarthy08}) photometric data for approximately 230 million distinct sources over an area of 8240 deg$^2$. Current and future wide and deep field surveys such as SWIRE \citep{mcmahon01,irwinlewis01,rowanrobinson08}, UKIDSS \citep{lawrence07}, CFHTLS \citep{cabanac07,mccracken08}, Pan-STARRS \citep{kaiser05}, LSST \citep{sweeney09}, VST ATLAS, the VISTA Hemisphere Survey and the SDSS itself will continue to add to the mapping of the Universe around us presenting increasing amounts of data at a variety of wavelengths. Given this enormous effort in the collection of photometric data, and the expense of subsequent spectroscopic surveys, the filtering of galaxies by type and redshift via their photometric properties is a vital and powerful tool for the effective use of the large quantities of photometric data available to us. Selecting distinct galaxy populations in this way offers a relatively cheap route to large galaxy and QSO surveys, either through using photometric redshifts or by using broader photometric constraints to select specific galaxy populations for subsequent spectroscopic surveys. 

A key example of the success of this process over recent years has been the photometric selection of Luminous Red Galaxies (LRGs) at redshifts up to z$\approx$0.8 \citep{eisenstein01, padmanabhan05a, blake, collister07}. LRGs offer a simple route to photometric selection of different redshift samples due to the 4000\AA\ break feature, which passes through the optical wavelength bands from redshifts of $z=0$ out to $z=1$. Such selections have been used in a range of key cosmological applications and perhaps most significant amongst these is the detection of the baryon acoustic oscillation (BAO) signature in analyses of large scale structure. This was measured by \cite{eisenstein05} using spectroscopic redshifts of $>40,000$ LRGs selected from the SDSS using photometric constraints based on the progression of the 4000\AA\ break. The measurement of the BAO signal should also be possible using suitably accurate photometric redshifts as the BAO galaxy clustering features are on scales of order 100h$^{-1}$Mpc \citep{eisenstein01, cole, eisenstein05}. In this vein \citet{sawangwit} used some 500 LRGs with spectroscopic redshifts obtained using AAOmega on the AAT to calibrate an $r-i:i-z$ colour-cut in terms of redshift and obtained a redshift distribution in the range $0.55<z<1.0$, with $\overset{\_}{z}=0.68\pm 0.1$. They go on to use a full photometric sample based on this selection, combined with two others with $\overset{\_}{z}\approx0.35$ and $\overset{\_}{z}\approx0.55$, to investigate the BAO signal. Using these photometric selections to measure the BAO signature in the angular correlation function, they show an agreement with the spectroscopic measurements reported by \citet{eisenstein05}.

A further example of the use of photometrically selected LRGs is the study of the Integrated Sachs Wolfe effect (ISW), in which CMB photons are subjected to shifts in energy as they pass through the gravitational potential wells of galaxies and clusters in an accelerating Universe \citep{sw}. This has been studied by a number of authors using the WMAP 1st and 3rd year data releases in combination with photometrically selected galaxy populations. Aside from the use of simple magnitude limited galaxy samples (e.g. \citealt{fosalba03,fosalba04,rassat07}), this work is again dominated by the use of LRGs. For example, \citet{padmanabhan05b} use photometrically selected LRGs from the SDSS (covering a redshift range of $0.2<z<0.6$) to make a $2.5\sigma$ detection of the ISW effect in WMAP 1st year data, whilst \citet{cabre06} combine a $z\approx0.5$ SDSS LRG sample with the WMAP 3rd year data to claim a detected ISW signal of $3\sigma$. More recently \citet{giannantonio08} performed a combined analysis using a number of galaxy and QSO catalogues, incorporating 2MASS data \citep{jarrett00}, SDSS DR6 galaxies, SDSS MegaZ LRGs \citep{collister07}, NVSS radio data \citep{condon98}, HEAO X-ray data \citep{boldt87} and the SDSS DR6 QSO catalogue \citep{richards09}. By combining the results from these datasets, \citet{giannantonio08} report a $4.5\sigma$ detection of the ISW effect and go on to use this to test a number of cosmological models. In particular, their ISW result places constraints on the mass density of $\Omega_m=0.26^{+0.09}_{-0.07}$, assuming a flat $\Lambda$CDM cosmology.

As an alternative to the recent dominance of LRGs for large redshift surveys, we now look at the photometric selection and clustering of emission line galaxies (ELGs). The key advantage of using ELGs in this application is the ability to identify galaxies through spectroscopic observations with relatively short exposure times, due to the emission lines with which they are most easily identified. \citet{wigglez} have successfully used this to their advantage to undertake the WiggleZ spectroscopic survey of ELGs with the aim of measuring the BAO signal at z$\approx$0.7. They select ELGs in the redshift range $0.5<z<1$ using a combination of GALEX FUV and NUV imaging with g, r and i band optical imaging from SDSS and the second Red Cluster Sequence (RCS2) project \citep{yee07}. In this case the GALEX data allows them to select $z>0.5$ galaxies using the Lyman Break technique, whilst the SDSS optical data allows them to limit their sample to just blue emission line objects. They then perform spectroscopic observations using the AAOmega instrument at the 4m AAT, with exposure times of $<1hr$ required for successful identification.

In this paper we attempt to refine a number of photometric selections of ELGs in different redshift ranges, using optical data from the SDSS. In making greater use of the ELG population in studies of large scale structure, we may maximise the use of the available data from such large scale surveys as the SDSS and the upcoming VST ATLAS and Pan-STARRS surveys. We use COMBO-17 photometric redshift data in combination with SDSS data to perform a calibration of our photometric selections (section~\ref{sec:selection}). We go on to outline and review our observation program at the AAT, which was aimed at providing a spectroscopic catalogue with which to further calibrate the photometric selections. In section~\ref{sec:clustering} we then evaluate the clustering properties of the galaxy populations contained in our photometric selections using SDSS data. We then use the full samples of $>$600,000 galaxies selected from the SDSS to perform a search for the ISW effect in WMAP 5yr data (section~\ref{sec:isw}). We present our summary and conclusions in section~\ref{sec:summary}. Throughout this paper we assume a $\lambda$CDM cosmology with $\Omega_M=0.3$, $\Omega_{\Lambda}=0.7$ and $H_0=100h\kps$Mpc.

\section{Photometric Selection}
\label{sec:selection}
\subsection{Data \& Selection}

Using SDSS imaging data,  our aim is to develop a set of photometric selection criteria using the SDSS filter bands \citep{fukugita96} alone to isolate emission line galaxies in 3 separate redshift ranges of approximately $z<0.4$, $0.4<z<0.6$ and $z>0.6$. At these redshifts, the 4000\AA\ break is a key feature in the observed optical spectra of both red and blue galaxies as it moves through the g and r SDSS filters with increasing redshift. In ELGs however, the break is somewhat weaker than in the spectra of LRGs, whilst the continuum at wavelengths greater than 4000\AA\ remains lower compared to the LRG spectrum due to the dominance of young blue stars in the ELGs. These contrasts in the spectra of LRGs and ELGs inherently allow us to separate the two in colour space, whilst simultaneously facilitating photometric selections of galaxies at different redshifts.

\begin{figure}
  \centering \includegraphics[width=80.mm]{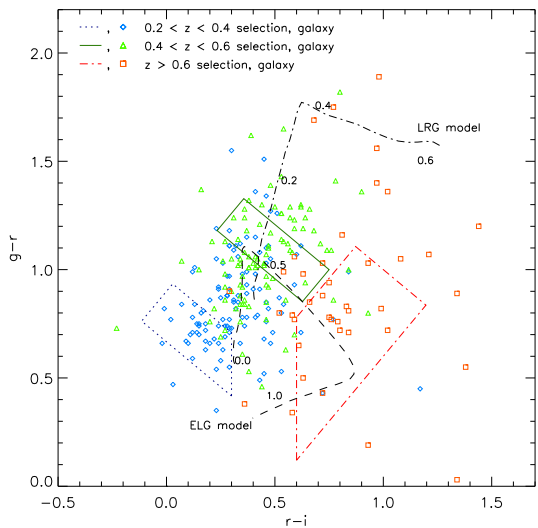}
  \caption[\small{GRI plot of COMBO17 galaxies.}]{\small{Galaxies in the COMBO-17 S11 field plotted in the gri colour plane using SDSS magnitudes. The cyan diamonds, green triangles and red squares show galaxies with photometric redshifts in the ranges $0.2<z<0.4$, $0.4<z<0.6$ and $z>0.6$ respectively. The labelled black tracks show the evolution of an emission line and an elliptical galaxy based on a Salpeter model, from $z = 1.2$ to $z = 0$. Our photometric selections are marked by the dotted blue, solid green and dash-double-dotted boxes for our low, mid and high redshift bins.}}
  \label{combocol}
\end{figure}

With this in mind we have used the Bruzual and Charlot stellar population synthesis code \citep{bc03} to model the evolution of a typical emission line galaxy in the gri (AB) colour plane. We used a Salpeter IMF with a galaxy formed at $z=6.2$ (i.e. with an age of 12.6Gyr at $z=0$) and a $\tau=9$Gyr exponential SFR. The resultant gri colour evolution track from $z=1.2$ to $z=0$ is shown in figure~\ref{combocol} (dashed black line). Here we see a clear evolution in the gri colour space around which we may build a selection regime for identifying candidates in our desired redshift ranges. We also plot a track (dot-dash line) for an elliptical galaxy using a $\tau=1$Gyr exponential SFR and a redshift of formation of z=10 (with Solar metallicity and Salpeter IMF).

\begin{table*}
\caption{Selection criteria chosen to identify galaxies in our three
redshift ranges: $0.2<z<0.4$, $0.4<z<0.6$ and $0.6<z<1.0$ using SDSS ugriz AB magnitudes. These are
illustrated in the ugr colour plane in figure~\ref{combocol}.}
\centering
{\small
\begin{tabular}{@{}ccc@{}}
\hline
Low redshift           & Mid-redshift                & High redshift               \\
\hline
$19.0<i<20.0$       & $19.0<i<20.2$        & $19.5<i<20.5$      \\
$r-i<0.3$           & $g-r>1.2(r-i)+0.1$   & $r-i>0.5$          \\
$g-r<1.2(r-i)+0.9$  & $g-r<1.2(r-i)+0.9$   & $g-r<1.2(r-i)+0.06$\\
$g-r>-1.2(r-i)+0.75$& $g-r>-1.2(r-i)+1.65$ & $g-r>1.2(r-i)-0.6$ \\
$g-r<-1.2(r-i)+1.3$ & $g-r<-1.2(r-i)+1.95$ & $g-r<-1.2(r-i)+2.2$\\
                    & $r-i>-(i-z)+0.5$     &                    \\
                    & $-2.0<u-g<1.0$       &                    \\
                    & $i - z > 0.55$       &                    \\
\hline
   \end{tabular}
}
\label{tab:sel}
\end{table*}

We calibrate our selections using the photometric redshift data published by the COMBO-17 team \citep{wolf2,simon08}. The data we use is from the COMBO-17 S11 field, which covers an area of $0.5^{\circ}\times0.5^{\circ}$ centred at 11h 42m 58s, -01 42$\arcmin$ 50$\arcsec$ (J2000) and the entirety of which is covered by SDSS imaging. These data provides accurate ($dz/(1+z)=0.02$) photometric redshifts for a total of 7248 galaxies based on broad and narrow band imaging using 17 different optical and near-infrared filters. We match the positions of COMBO-17 galaxies to the equivalent objects in the SDSS data, thus combining the SDSS $ugriz$ magnitudes with the COMBO-17 photometric redshifts. The S11 objects are shown in the (SDSS) $g-r$, $r-i$ colour plane in figure~\ref{combocol}. For the purposes of clarity, we only plot those galaxies classed as blue spirals by the COMBO-17 team in this figure, however the location of red-sequence galaxies is indicated by the LRG evolution track (dot-dash line). The COMBO-17 galaxies have been split into three populations for the purposes of this plot based on their assigned photometric redshift from COMBO-17: $0.2<z<0.4$ (blue diamonds), $0.4<z<0.6$ (green triangles) and $0.6<z<1.0$ (red squares).

Based on the distribution of the photometric redshifts and the ELG evolution track presented in the above plot, there is a clear progression in the gri colour plane based on ELG redshift. Further to this, areas of the plot can be isolated that should minimize the number of red-sequence galaxies, whilst maximizing the numbers of either $z<0.4$ or $z>0.6$ ELGs. The medium redshift range does however present significant problems. The ELG evolution track appears to pass through a region populated by both lower and higher redshift ELGs as well as low-redshift red-sequence galaxies in the $0.4<z<0.6$ range.

From the above observations we construct three sets of colour cuts to preferentially select three redshift ranges. These are shown in figure~\ref{combocol} by the dotted blue box (low redshift cut), solid green box (medium redshift cut) and dash-double-dot red box (high redshift cut). As discussed above, the mid redshift range is significantly exposed to contamination from both ELGs at unwanted redshifts and red-sequence galaxies. To minimize the numbers of these we have therefore added colour cuts to this sample based on the $r-i$, $i-z$ and $u-g$ colours of the selected galaxies. These additional cuts have also been calibrated using the COMBO-17 photometric redshifts. The details of our three selections, including the additional mid-redshift colour cuts, are given explicitly in table~\ref{tab:sel}. These cuts have been tailored to produce sky densities of candidates of $\approx100deg^{-2}$ for each of the three redshift ranges in order to provide candidate numbers suitable for wide field spectroscopic surveys performed with instruments such as the 2dF/AAOmega spectrograph. 

\begin{figure}
  \centering \includegraphics[width=80.mm]{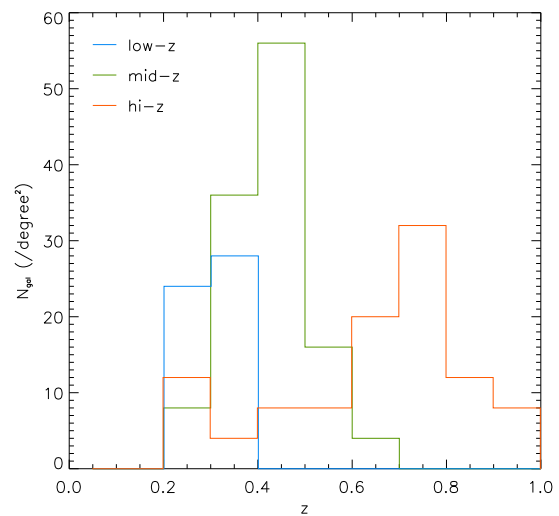}
  \caption[\small{}]{\small{Redshift distributions of our three
  photometric selections based on photometric redshifts from the
  COMBO-17 data in the S11 field. The three samples give mean
  redshifts of $\overset{\_}{z}_{lo}=0.29\pm0.05$,
  $\overset{\_}{z}_{mid}=0.44\pm0.08$ and
  $\overset{\_}{z}_{hi}=0.65\pm0.21$.}}
  \label{photz}
\end{figure}

The photometric redshift distribution for our three samples are shown in figure~\ref{photz}. This plot includes all the selected galaxies from the S11 field, including those identified as being part of the red-sequence (these making up $\approx4\%$ of the total selected across all three selections). The three selections are characterised by mean redshifts of $\overset{\_}{z}_{low}=0.29$, $\overset{\_}{z}_{mid}=0.44$ and $\overset{\_}{z}_{hi}=0.65$, with standard deviations of $\sigma_{low}=0.05$, $\sigma_{low}=0.08$ and $\sigma_{hi}=0.21$.

For the purposes of this paper we now use our three selections to create three datasets from the SDSS galaxy catalogue. We apply the selections to the SDSS DR6 data release taking our data from the PhotObjAll table. Aside from the colour-magnitude criteria given in table~\ref{tab:sel}, we reject objects which do not meet the following criteria:
\begin{itemize}
{\small
\item{$TYPE=3$ (i.e. classed as a galaxy);}
\item{$NCHILD=0$;}
\item{Flagged as BINNED1, BINNED2 and BINNED3;}
\item{$90^\circ<R.A.<270^\circ$.}
}
\end{itemize}
We also limit our selection to the main SDSS region, rejecting stripes 40-43. Stripe 26 is also rejected as this appears to show some contamination and spurious density fluctuations. The photometric selections are performed using the SDSS model magnitudes with the appropriate extinction values subtracted. The total numbers of candidates given by each selection are 892,528, 620,020 and 734,566 for the low-, mid- and high redshift selections respectively. These numbers give sky densities of $103deg^{-2}$, $71.9deg^{-2}$ and $85.1deg^{-2}$.

\subsection{Observations}

An important element of this work is the calibration of the photometric selection samples with spectroscopic observations to confirm the achievable redshift distribution of our selections. To this end, we have performed spectroscopic observations of our $z\approx0.7$ sample using the AAOmega spectrograph at the AAT \citep{saunders04,sharp06}. AAOmega is a double beam spectrograph fed by $2.1\arcsec$ diameter fibres, which allows the simultaneous observations of up to $\approx360$ objects in a circular field of view of diameter $2^\circ$.

Observations were taken on the AAOmega instrument at the Anglo-Australian Observatory (AAO) on the nights of 4th and 6th March 2006. The spectrograph was configured using the 5700\AA\ dichroic, with the 580V grating mounted in the blue-arm and the 385R grating in the red-arm. The 580V grating gives a wavelength coverage of 370nm to 580nm, with a pixel size of 0.1nm/pixel and the 385R a coverage of 560nm to 880nm, with a pixel size of 0.16nm/pixel. Both provide a resolution of 1300. In total, AAOmega offers 400 fibres per observation, however a significant number of these were at times used for other projects (e.g. \citealt{ross2}), were locked to guide-stars, sky-targets or simply malfunctioning and so our target numbers range from $\sim$50-230 per observation. We targeted four 2$^{\circ}$ fields with multiple exposures of 1200s each. The observations are summarised in table~\ref{observations}. The four observed fields are targeted towards the Cosmic Evolution Survey (COSMOS, \citealt{scoville07}) field, the d05 and e04 fields from the 2dF-SDSS LRG and QSO (2SLAQ, \citealt{croom04}) survey and the S11 field from COMBO-17.

\begin{table}
\caption{Co-ordinates of the four fields targeted with the number of
gri selected ELG candidates in each 2-degree field.}  \centering
  \begin{tabular}{@{}lcccc@{}}
  \hline 
  Field & COSMOS & d05 & e04 & S11      \\
  \hline 
  Date        & 06/03/06  & 04/03/06  & 04/03/06  & 06/03/06 \\
  R.A.          & 150.118$\deg$& 200.399$\deg$    & 221.899$\deg$      & 175.741$\deg$  \\
  Declination   & +2.2052$\deg$& -0.2124$\deg$    & -0.2141$\deg$      & -1.7159$\deg$ \\
  Exposure time & 3$\times$1200s& 3$\times$1200s& 4$\times$1200s&3$\times$1200s\\
   Seeing        & $\approx3.0\arcsec$& $\approx2.0\arcsec$& $\approx2.5\arcsec$&$\approx1.6\arcsec$\\
  Candidates    & 378    & 329        & 343          & 391      \\
  Targeted      & 217    & 45   & 225 & 219 \\
  ELG redshifts & 44 & 10 & 84 &  142 \\ \hline
\end{tabular}
\label{observations}
\end{table}

Target objects were selected using our selection criteria applied to the SDSS data available for each of the fields. Over the course of the 5 nights, seeing ranged from $\sim1.5\arcsec-3.0\arcsec$, with a mean of $\sim2.0\arcsec$. All observations were flat fielded, arc calibrated and combined using the AAO's 2dFDR tool. Approximately 20$\%$ of fibres were affected by an early instrumentation problem known as fringing, which led to an almost sinusoidal signal in the output. In the S11 field this affected 27 of the fibres targeted on ELG candidates. A further problem, is encountered due to the strong sky emission lines above 8250\AA. These limit the identification of H$\beta$ and [OIII] above $z\approx0.65-0.7$, however they do not interrupt identification of [OII] and so the impact of the sky lines is limited.

\begin{figure*}
 \centering \includegraphics[width=180.mm]{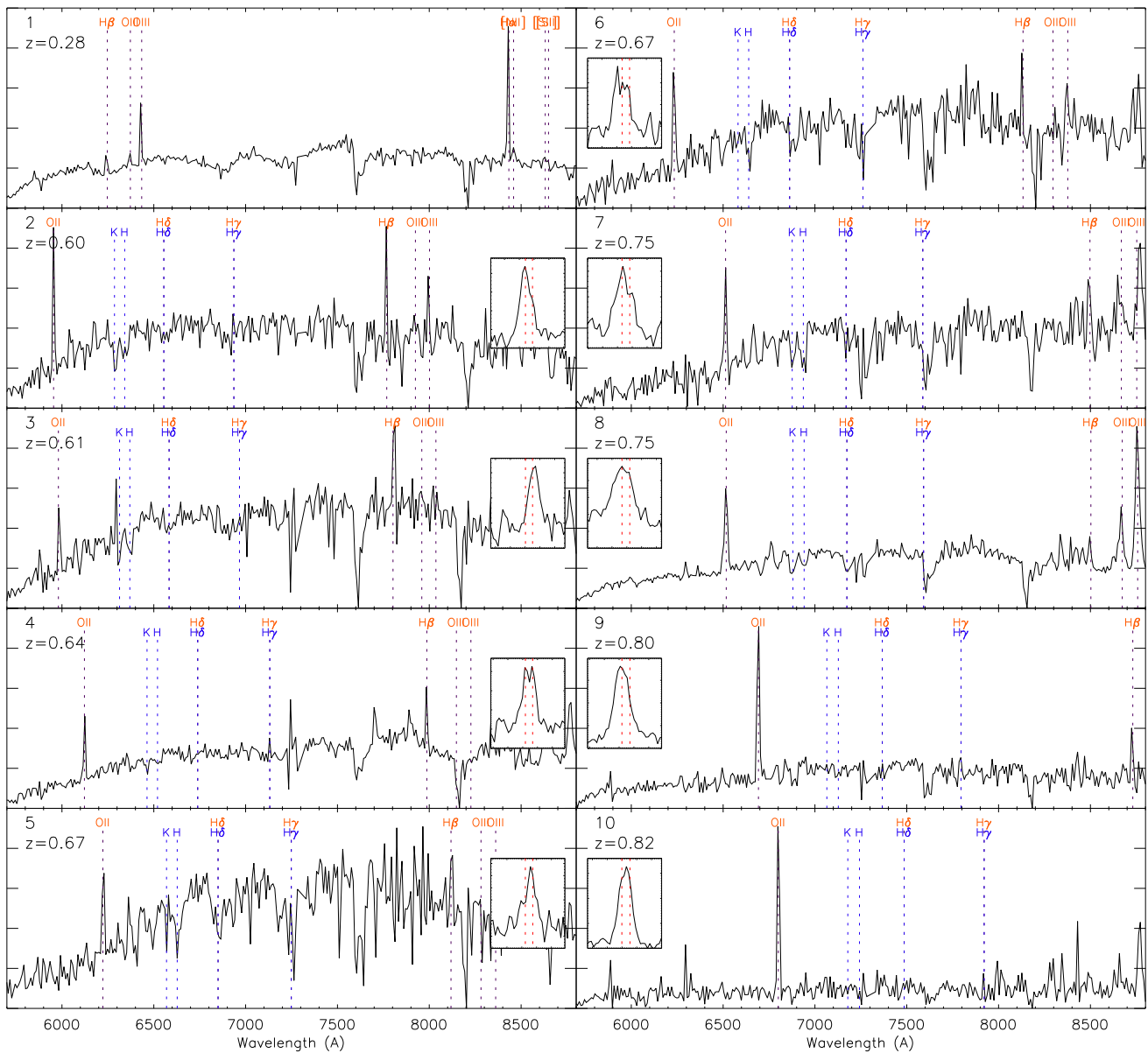}
 \caption{Example spectra taken on the AAOmega spectrograph with the 385R grism, binned to 10\AA\ bins. Wavelengths of galaxy emission and absorption features are marked, however the features of key use in identification were the [OII], H$\beta$ and [OIII] doublet emission lines. The insets each show the [OII] feature in close-up and unbinned, showing the doublet nature of the [OII] feature to be marginally discernible given the resolution of the spectrograph. The red dashed-lines show the expected positions of the doublet peaks at rest-wavelengths of 3726\AA\ and 3729\AA. Objects 1, 3, 5 and 6 are all marked in figure~\ref{photoz} as showing discrepancy between spectroscopic and photometric redshifts.}
 \label{elgspec}
\end{figure*}

\subsection{Galaxy redshifts}

We use the 2dFGRS software {\small AUTOZ} \citep{croom01} and {\small RUNZ} \citep{colless01} to search for emission features with which to identify galaxies in our observed sample and determine redshifts. {\small AUTOZ} performs an initial identification of each spectrum, fitting absorption and emission features. Each fibre spectrum was then evaluated by eye to assign a redshift and quality rating, $q_{op}$ (which ranges from 0 to 5 depending on the confidence of the identification). Only objects with $q_{op}\geq3$ were accepted as positive identifications.

Examples of the spectra obtained with the AAOmega instrument are provided in figure~\ref{elgspec}. The spectra are all binned to a bin-width of $\approx10$\AA\ and key emission and absorption features are marked. We also show the unbinned data for the [OII] feature (insets), where it is evident that the doublet nature of the feature is marginally detectable at the observed resolution. Although data were obtained on both the blue and red arms, only spectra from the red arm are plotted here as there are few features useful for identification in the blue wavelength range given the signal to noise of our data. The key emission features that facilitate the identification of these galaxies with short exposure times, i.e. [OII], H$\beta$ and the [OIII] doublet are all evident in these spectra.

There were no positive identifications of galaxies with just absorption line features and no emission lines in the spectroscopic observations. It is likely however that any absorption line galaxies targeted remained unidentified given the relatively short exposure times, which make it difficult to confidently identify absorption line features in the spectra.

A summary of the numbers of emission line galaxies identified in our four fields is provided in table~\ref{observations}. Our most successful field was the COMBO-17 S11 field in which we were able to target 219 ELG candidates in seeing conditions of $\approx1.6\arcsec$. In this field we identified 121 of the 219 candidates as being emission line galaxies from their emission lines with a confidence of $q_{op}\geq3$. In total we were able to identify 311 emission line galaxies over a combined area of 12.4 deg$^2$, giving an average sky density of 25 deg$^{-2}$. However, three of the observed fields suffered poor seeing conditions of $\geq2.0\arcsec$, limiting our ability to successfully identify objects in these fields. At worst, completeness was reduced to $<25\%$ in the COSMOS field due to the seeing of $\approx3.0\arcsec$. However, in the more reasonable observing conditions encountered with the observations in the S11 field (where the seeing was $1.6\arcsec$) we find that the identification rate is a more promising $\sim65\%$, with a sky density of $\approx40deg^{-2}$.

\begin{figure}
\includegraphics[width=80.mm]{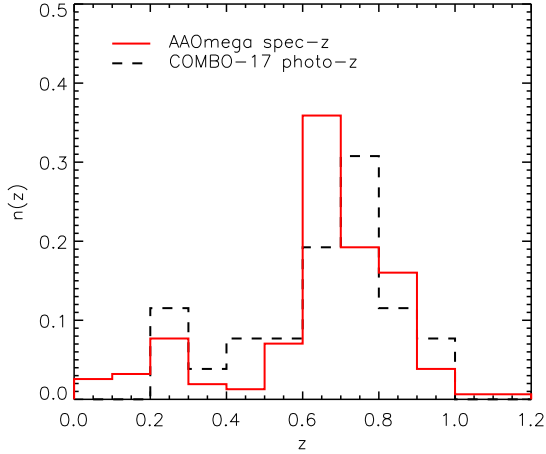}
  \caption{The redshift distribution of all successfully identified emission line galaxies from the four fields targeted using AAOmega is shown (red histogram). A total of 280 galaxies were successfully identified. The original photo-z redshift distribution incorporating all the galaxies selected using our $z\approx0.7$ colour cuts in the COMBO-17 S11 field is also shown (black dashed line). Both distributions are normalized to give an area under the histogram equal to unity.}
  \label{nz}
\end{figure}

Figure~\ref{nz} gives the redshift distribution of the spectroscopically confirmed galaxies in the S11 field. The plot incorporates all galaxies identified in the S11 field and the original photometric redshift distribution from COMBO-17 data (black dashed line) also from the S11 field. Our spectroscopic sample follows the expected distribution closely, with $\overset{\_}{z}_{spec}=0.66\pm0.23$ (compared to $\overset{\_}{z}_{phot}=0.65\pm0.21$). There is some contamination from lower redshift (i.e. $z<0.5$) galaxies and in the spectroscopic sample this is at a level of $\approx18\%$ (compared to a level of $\approx23\%$ obtained with the COMBO-17 photometric redshift sample).

\begin{figure}
  \centering \includegraphics[width=80.mm]{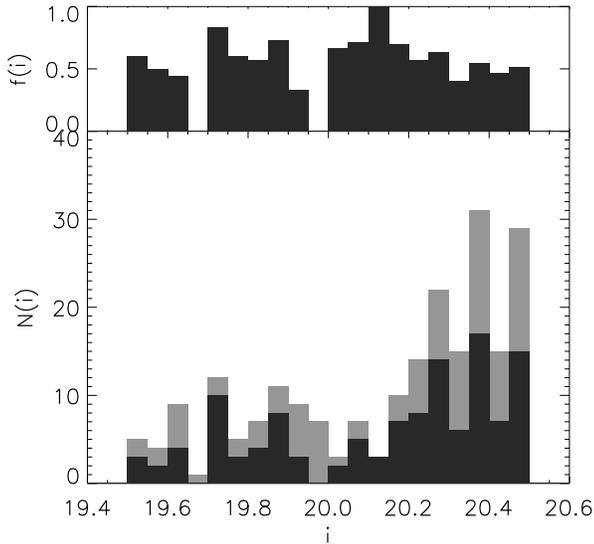}
  \caption[\small{}]{\small{Number counts of objects observed as a
  function of SDSS i-band magnitude. The dark histogram shows counts
  of objects identified with emission lines ($N_{em}$), whilst the
  pale histogram shows the total number of objects ($N_{T}$) observed
  in each magnitude bin. The top panel shows the fraction
  $N_{em}$/$N_{T}$ as a function of i-band magnitude. Data are only
  shown from the S11 field as all other fields were limited by adverse
  seeing.}}
  \label{magcom}
\end{figure}

Figure~\ref{magcom} shows identifications as a function of source magnitude in the S11 field. The lower panel shows number counts of spectroscopically confirmed galaxies exhibiting emission lines ($N_{em}$, dark bars) and of all objects targeted with AAOmega fibres ($N_{T}$, pale bars), whilst the upper panel shows the fraction, $N_{em}$/$N_{T}$. The consistency of the 65\% identification rate across our magnitude range is evident and we are clearly reaching the $i=20.5$ magnitude limit successfully. A small fall-off in the fraction of ELGs identified is observed in the fainter magnitude bins, however numbers still remain high.

\begin{figure}

 \includegraphics[width=80.mm]{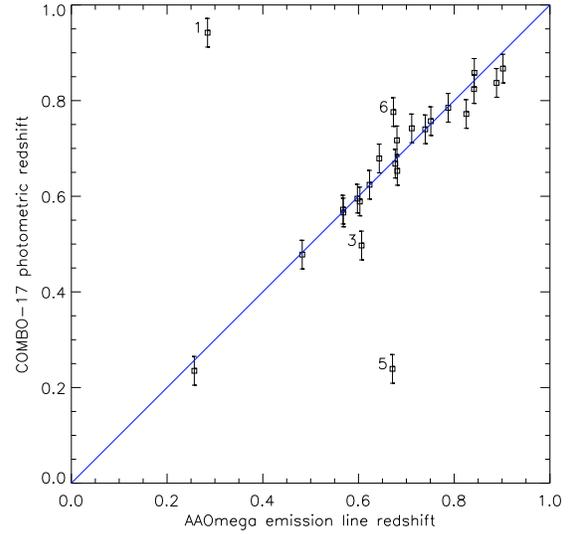}
 \caption{Comparison of the spectroscopic redshifts of 24 galaxies in the central $0.5^\circ\times0.5^\circ$ of the S11 field with photometric redshifts from the COMBO-17 survey. The error bars show the
$\sigma_z\sim0.03$ error quoted by the COMBO-17 team for their photometric redshifts. The four numbered points (1, 3, 5 and 6) are objects in which the photometric and spectroscopic redshifts disagree by more than $3\sigma_z$ (the numbers refer to the spectrum numbers from figure~\ref{elgspec}). }
 \label{photoz}
\end{figure}

In figure~\ref{photoz}, we compare our spectroscopically determined redshifts against the COMBO-17 photometric redshifts for those galaxies lying in the central $0.5^{\circ}\times0.5^{\circ}$ region covered by the COMBO-17 data. The vertical error bars represent the $\sigma_z\sim0.03$ error quoted by the COMBO-17 team for the photometric redshifts. We find a total of 24 objects that have both COMBO-17 photometric redshifts and spectroscopic redshifts from this work. Overall there appears to be good agreement between the data with just 4 outliers (taken here as a difference between the photometric and spectroscopic results of $3\sigma_z$) having significantly different redshifts. The spectra for all four of these objects are given in figure~\ref{elgspec} and each of the outliers are marked in figure~\ref{photoz} by the spectrum number (1, 3, 5 and 6) from figure~\ref{elgspec}. We find a mean offset between the spectroscopic and photometric redshifts of $\Delta z = 0.01\pm0.04$ (after excluding points 1 and 5).

\begin{figure}
 \includegraphics[width=80.mm]{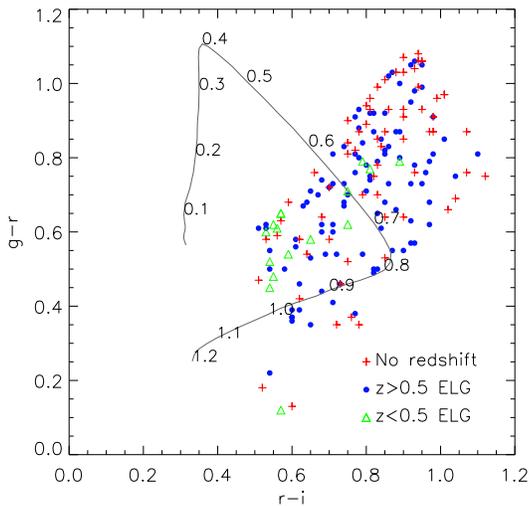}
 \caption{Spectroscopic results from the S11 and e04 fields. We show objects identified as $z\geq0.5$ emission line galaxies (filled blue circles), $z\leq0.5$ emission line galaxies (green triangles) and objects with no identified redshift (red crosses). The same evolution track as plotted in figure~\ref{combocol} is also shown.}
 \label{colours}
\end{figure}

We show in figure~\ref{colours} the distribution of spectroscopically confirmed $z>0.5$ ELGs (filled blue cicles), $z<0.5$ ELGs (green triangles) and unidentified objects (red crosses) in the g-r vs r-i colour plane. It is evident that the $z>0.5$ emission line galaxies are reasonably evenly spread in the g-r vs r-i colour plane as are the objects without any discernible emission, although there is some bias in these to be towards the redder end of the selection in both $r-i$ and $g-r$. The $z<0.5$ emission line galaxies appear to be biased towards the upper left limits of the selection region, towards the low-redshift main sequence. These may be further reduced by altering our constraints, however this would also remove a significant number of $z>0.5$ objects at the same time. The model evolution track from figure~\ref{combocol} is again plotted for reference.

\begin{figure}
 \includegraphics[width=80.mm]{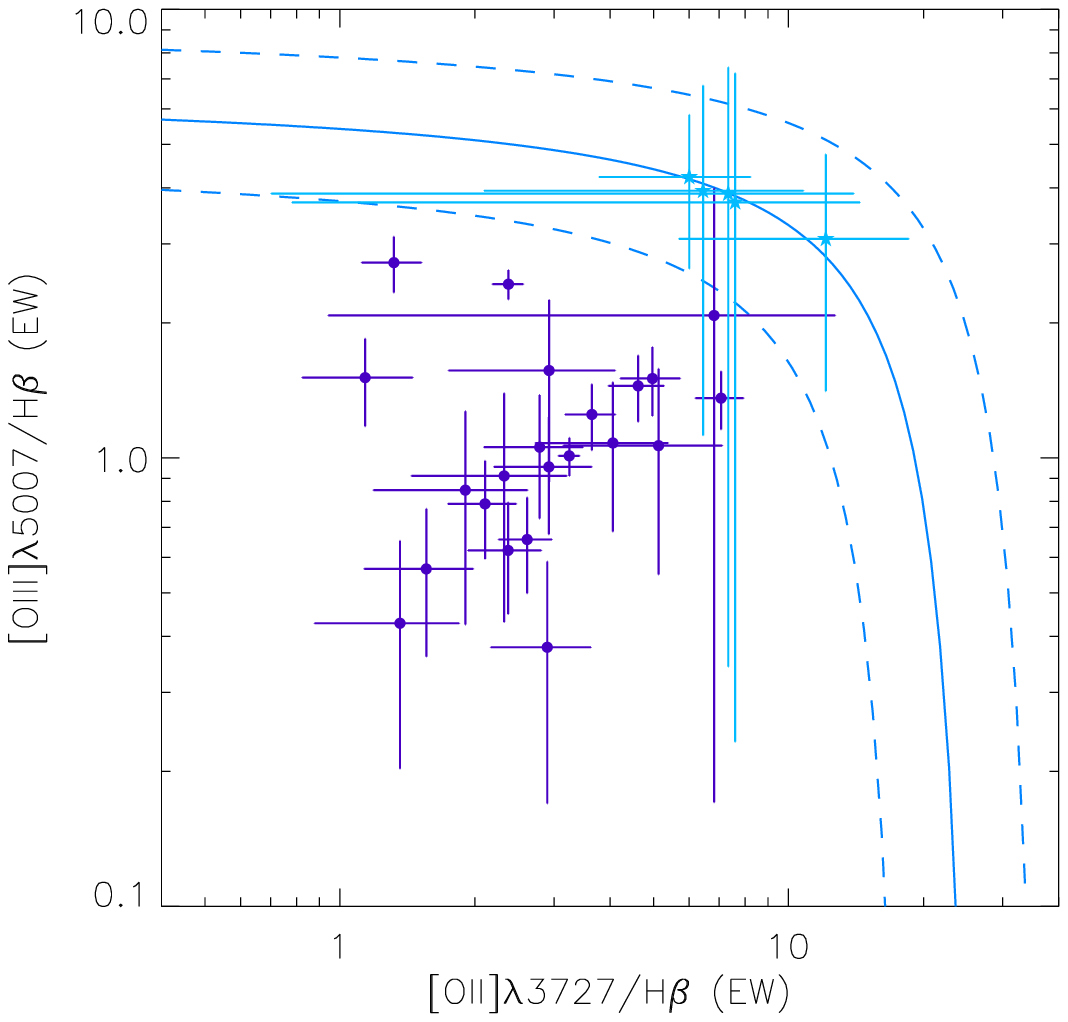}
 \caption{"Blue diagnostic" diagram based on \citet{lamareille04}. Line ratios are plotted for the sub-sample of our spectroscopically observed sample for which we have equivalent widths for the [OII]$\lambda3727$, H$\beta$ and [OIII]$\lambda5007$ nebular emission lines. The solid line marks the limit estimated by \citet{lamareille04} between star-forming galaxies and AGN and the dashed lines show the region of uncertainty. In total 22 objects lie within the star-forming region of the diagnostic plot and are marked by filled blue circles. A further 5 lie within the overlap region (cyan stars) and none of the objects lie within the AGN region.}
 \label{agn}
\end{figure}

Now looking at the properties of the galaxy spectra, we measure the equivalent widths of the nebular emission lines by fitting Gaussian curves to the [OII] 3727\AA, H$\beta$ and [OIII] 5007\AA\ lines. We were able to measure equivalent widths with confidence for [OII] 3727\AA, H$\beta$ and [OIII] 5007\AA\ in 109, 53 and 51 of the galaxies in our sample respectively. From these we determined mean equivalent widths of 23.0\AA, 8.12\AA\ and 8.98\AA\ for [OII] 3727\AA, H$\beta$ and [OIII] 5007\AA\ respectively. These mean equivalent widths are broadly consistent with other measurements of emission lines in late-type galaxies (e.g. \citealt{kenn92a,shi06}). In 27 of these galaxies we were able to measure all three of the above nebular emission lines with confidence and have attempted to evaluate the presence of AGN in our sample using the "blue diagnostic" constraints of \citet{lamareille04}, which are based on the [OII]$\lambda3727/$H$\beta$ and [OIII]$\lambda5007/$H$\beta$ line ratios. This is shown in figure~\ref{agn}, where the solid line marks the estimated division between AGN (above) and star-forming galaxies (below). The dashed lines mark the region of uncertainty between the two populations. In all, 22 of this sub-sample fall within the star-forming galaxy region, whilst the remaining 5 (2 of which have large uncertainties) fall within the uncertain region and none lie in the AGN region. Within the reliability of the blue diagnostic diagram, we can say that our sample is dominated by star-forming galaxies and this method shows no positive evidence for any AGN contamination of our sample although there are a small number of borderline cases.

\begin{figure*}
 \centering \includegraphics[width=180.mm]{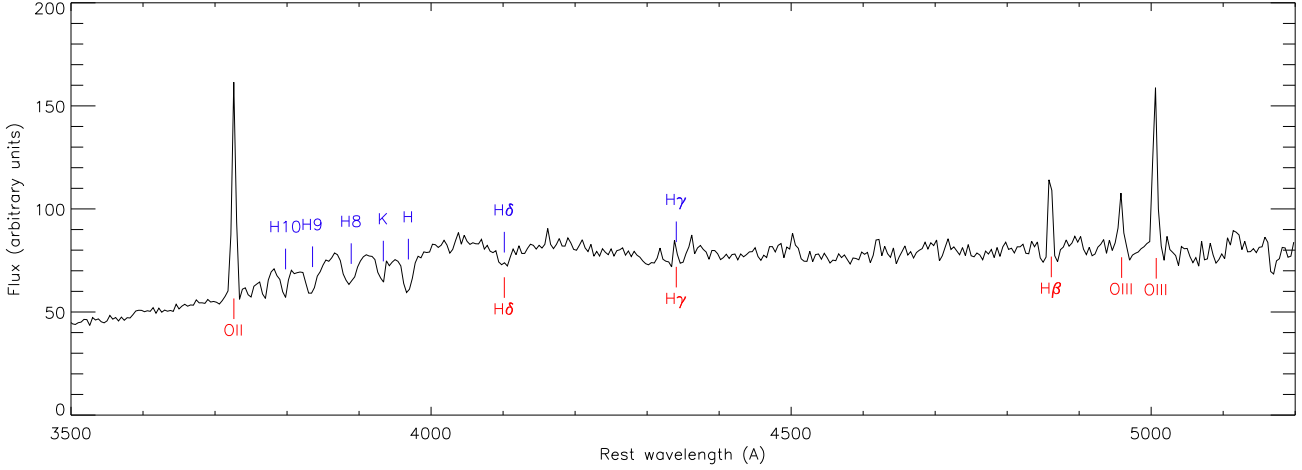}
 \caption{Composite spectrum of the 280 successfully identified  emission line galaxies. The key emission line features used for identification are clearly visible: [OII], H$\beta$ and [OIII], whilst  absorption features which are difficult to observe in individual spectra are now evident.}
 \label{composite}
\end{figure*}

Figure~\ref{composite} shows a composite spectrum of all of the confirmed emission line galaxies over all redshifts, with significant emission and absorption features labelled. The key emission lines used in our spectral identification (i.e. [OII], H$\beta$ and [OIII]) are clearly evident. We also see the Balmer absorption features red-wards of the [OII] emission, whilst the weak ELG 4000\AA\ break is also apparent in this composite.

\section{Clustering}
\label{sec:clustering}
\subsection{Angular Correlation Function}

\begin{figure}
 \includegraphics[width=80.mm]{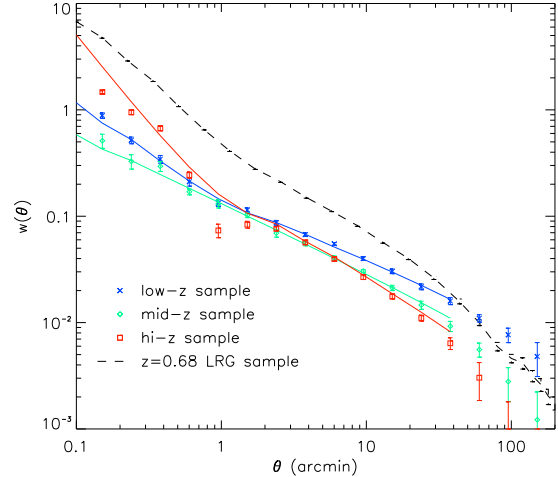}
 \caption{\small{The angular correlation functions, $w(\theta)$ for our three photometric redshift selections. Blue crosses, green diamonds and red squares represent the low- ($z<0.4$), mid- ($0.4<z<0.6$) and high-redshift ($z>0.6$) samples respectively. The best fitting power law models are plotted through each set of data. The dashed black line shows the $w(\theta)$ of \citet{sawangwit} for their sample of photometrically selected Luminous Red Galaxies at $z=0.68\pm0.1$.}}
 \label{wtheta}
\end{figure}

We now evaluate the angular correlation function for a sample of galaxies selected based on our three photometric selections. The datasets taken from SDSS DR6, as described in the previous section, are used for this purpose. We calculate the angular-correlation function of the samples using the
estimator:

\begin{equation}
\label{eq:wth}
w(\theta) = \frac{n_R}{n_D}\frac{DD}{DR} - 1
\end{equation}

\noindent where $DD$ and $n_D$ are the numbers of galaxy-galaxy pairs and the total number of galaxies respectively. For these calculations we use a random catalogue which exactly matches the sky coverage of our SDSS galaxy samples and with a factor of 20 more random points than galaxies in each of our galaxy samples. The total number of random points is given by $n_R$ and $DR$ is simply the number of galaxy-random pairs. Statistical errors are estimated using field-to-field errors, using 16 separate fields within our complete field. Our results for the three photometric samples are shown in figure~\ref{wtheta} where the blue triangles, green squares and red crosses show the low, mid and high redshift samples respectively. For comparison we also plot $w(\theta)$ for a photometrically selected LRG sample from SDSS data at $z\approx0.7$ \citep{sawangwit}. The clustering amplitude of all three of our samples is significantly lower than the LRG amplitude.

From these measurements of the angular correlation function, we now estimate the 2-point correlation function (2PCF, $\xi(r)$) using Limber's formula. We make an estimate of $\xi(r)$ for each of the samples using a double power-law with a central break, i.e. $\xi(<r_b)=(\frac{r}{r_{0,1}})^{-\gamma_1}$ and $\xi(>r_b)=(\frac{r}{r_{0,2}})^{-\gamma_2}$. This is then combined with our best estimate of the redshift distribution (based on the COMBO-17 photometric redshift data for the low and mid redshift samples and the spectroscopic redshift data for the high redshift sample) to calculate the resultant $w(\theta)$ with Limber's formula. A full treatment of this calculation is given by \citet{phillipps}. We then perform a $\chi^2$ fitting, in the range $2\arcmin<\theta<20\arcmin$ to our data. The best fitting models are plotted with the data in figure~\ref{wtheta}, whilst the associated parameters are listed in table~\ref{r0par}. We find reasonable fits to both the low- and mid-redshift samples, the low-redshift sample being well fitted by a double power law with a break at $0.5h^{-1}Mpc$ and the mid-redshift sample by just a single power-law. We note however, that we struggle to fit to the high-redshift sample with either a double or single power law. This is largely due to strong deviations from a simple power law trend at separations of $<2\arcmin$. This results in large $\chi^2$ values for our attempts to fit the correlation function at in this range. The angular correlation function does however return to a simple power law at separations of $2\arcmin<\theta<20\arcmin$ where we are able to provide a reasonable power-law fit using the Limber method.

\begin{table*}
\caption{Comoving correlation lengths, $r_0$ and power-law slopes, $\gamma$, for the double power-law model used to provide fits to the angular correlation functions for each redshift selection.}  \label{r0par}
{\small
\begin{tabular}{ccccccc}
\hline
z&$r_b$ ($h^{-1}$Mpc)&$r_0(<r_b)$ ($h^{-1}$Mpc)&$\gamma(<r_b)$&$r_0(>r_b)$ ($h^{-1}$Mpc)&$\gamma(>r_b)$&b\\
\hline
$0.29\pm0.05$&0.5&$1.30_{-0.03}^{+0.02}$&$2.21_{-0.01}^{+0.02}$&$2.65_{-0.08}^{+0.05}$&$1.54_{-0.02}^{+0.02}$&0.70\\
$0.44\pm0.08$&n/a&n/a&n/a&$3.62_{-0.06}^{+0.06}$&$1.65_{-0.01}^{+0.01}$&0.92\\
$0.65\pm0.21$&0.5&$2.30_{-0.05}^{+0.05}$&$2.72_{-0.01}^{+0.01}$&$5.88_{-0.12}^{+0.12}$&$1.83_{-0.01}^{+0.03}$&1.46\\
\hline
\end{tabular}
}
\end{table*}

From our estimates of $\xi(r)$, we now go on to estimate the bias of each sample. The biasing parameter, $b$, quantifies the relative clustering of a given galaxy population compared to the underlying dark matter distribution \citep{tegmark98}. This can be expressed as the following:

\begin{equation}
\xi_{gal}(r) = b^2\xi_{DM}(r)
\end{equation} 

Here, $\xi_{gal}(r)$ is the 2PCF of the galaxy sample and $\xi_{DM}(r)$ is the 2PCF of dark matter at the same epoch. We determine the dark matter correlation function by first using the {\small CAMB} (Code for Anisotropies in the Microwave Background, \citealt{lewis1}) software to estimate the DM power spectrum at the mean redshifts of each of our galaxy samples. The power spectrum is calculated using the HALOFIT model \citep{smith03} to fit non-linear features, at each of the mean redshifts of our samples. With the DM power spectra calculated at each redshift we then simply calculate the corresponding 2PCFs via the Fourier transform:

\begin{equation}
\label{psft}
\xi_{DM}(r) = 4\pi\int_0^\infty k^2P(k)\frac{sin(kr)}{kr}dk
\end{equation}

We now estimate the bias by evaluating the dark matter and the galaxy 2PCFs to a maximum separation of 20Mpc \citep{croom05, sawangwit}. This limit, restricts the calculations to the linear regime at which our fits to the correlation function are still valid. Thus the biasing parameter can be estimated using:

\begin{equation}
b^2 = \frac{\overset{\_}{\xi}_{gal}(20)}{\overset{\_}{\xi}_{DM}(20)}
\end{equation}

\noindent where $\overset{\_}{\xi}(x)$ is given by:

\begin{equation}
\overset{\_}{\xi}(x) = \int_{0}^{x} r^2\xi(r) dr
\label{eq:xi20}
\end{equation}

\begin{figure}
 \includegraphics[width=80.mm]{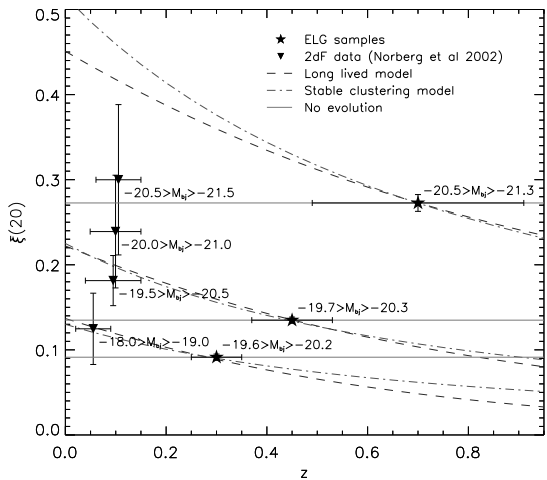}
\caption[{\small Estimated $\xi(20)$ plotted versus redshift for each of our three photometric samples.}]{\small Estimated $\overset{\_}{\xi}(20)$ plotted versus redshift for each of our three photometric samples (stars). Also plotted is the $\overset{\_}{\xi}(20)$ for late type galaxies from the 2dFGRS from \citet{norberg}, split by absolute magnitude (triangles). The dashed lines show the long-lived model normalised to the ELG data-points, whilst the dash-dot lines show the stable clustering model also normalized to the ELG data. The solid lines project the clustering of each population with no evolution.}
\label{xi20}
\end{figure}

We show $\overset{\_}{\xi}(20)$ for each of our three redshift samples in figure~\ref{xi20} (denoted by the stars). For comparison we also plot $\overset{\_}{\xi}(20)$ for the 2dFGRS late type galaxy samples of \citet{norberg}. These are split into absolute magnitude bins of $-18<M_{bj}-5log_{10}(h)<-19$, $-19<M_{bj}-5log_{10}(h)<-20$, $-20<M_{bj}-5log_{10}(h)<-21$ and $-20.5<M_{bj}-5log_{10}(h)<-21.5$ and are calculated based on the correlation parameters given in their table~\ref{r0par}. Based on the redshift and apparent magnitude distributions of our three samples, we estimate absolute magnitude ranges of $M_{b_j}-5log_{10}(h)=-19.9\pm0.3$, $-20.0\pm0.3$ and $-20.9\pm0.4$ for the low, mid and high redshift samples respectively. These estimates include K$+$e corrections based on the $\tau=9$Gyr SFR model given in section~\ref{sec:selection}.

Also plotted in figure~\ref{xi20} are three simple clustering evolution models: the long-lived model (dashed lines), stable clustering (dot-dashed lines) and no evolution of the comoving-space clustering (solid lines). All three models have been normalised to each of the ELG clustering amplitudes.

The long-lived model is equivalent to assuming that the galaxies have ages of order the Hubble time. The clustering evolution is then governed by their motion within the gravitational potential \citep{fry,croom05}. The bias evolution is thus governed by:

\begin{equation}
b(z) = 1+\frac{b(0)-1}{D(z)}
\end{equation}

\noindent where $D(z)$ is the linear growth rate and is determined using the fitting formulae of \citet{cpt92}. For the bias, b, we use the values given in table~\ref{r0par}: 0.70, 0.92 and 1.46 for the low, mid and high redshift samples respectively. The stable clustering model represents the evolution of virialised structures and is characterised by \citep{peacock}:

\begin{equation}
\overset{\_}{\xi}(r,z) \propto r^{-\gamma} \propto (1+z)^{\gamma-3}
\end{equation}

\noindent where $r$ is the comoving distance and $\gamma$ is the slope of the clustering correlation function. Finally, the no evolution model simply assumes that there is no evolution of the clustering in comoving coordinates.

Before analysing this data, it is important to note that the various samples have not been refined to be precisely equivalent in that neither the number densities or absolute magnitudes have been ideally matched between any of the samples. Further to this, the clustering strengths of the $z\approx0.3$ and $z\approx0.45$ are based on the photometric redshift distributions from the COMBO-17 S11 field, which as yet have not been fully calibrated. This comparison is intended to give only an indication of the potential clustering evolution of the the samples discussed.

Looking at the low and mid-redshift ranges we find that the clustering measurements suggest an increase in the clustering amplitudes of the emission line galaxies from redshifts of $z=0.3-0.5$ to more recent epochs if they are to evolve into the 2dF samples with comparable absolute magnitude ranges (i.e. the $-19.5>M_b>-20.5$ 2dF sample).

The high-redshift sample appears more consistent with a no evolution model if it is to evolve into the most comparable 2dF samples (i.e. $-20.5>M_b>-21.5$). However, the stable and long-lived clustering evolution models are still only $1-1.5\sigma$ above the brightest 2dF $\xi(20)$ value  and therefore cannot be ruled out by the present accuracy of the 2dF late-type data.

\subsection{Redshift-space Correlation Function}

We now estimate the redshift-space correlation function, $\xi(s)$, using $z\geq0.5$ galaxies identified with $q_{op}\geq3$ from the four fields observed with AAOmega. The redshift distribution from figure~\ref{nz} was used to create random catalogues with which to perform the auto-correlation analysis. In each field we use a catalogue of $20\times$ the number of random points as galaxies in that field. In total this calculation encompasses 276 galaxies across $12.6deg^2$. 

\begin{figure}
 \includegraphics[width=80.mm]{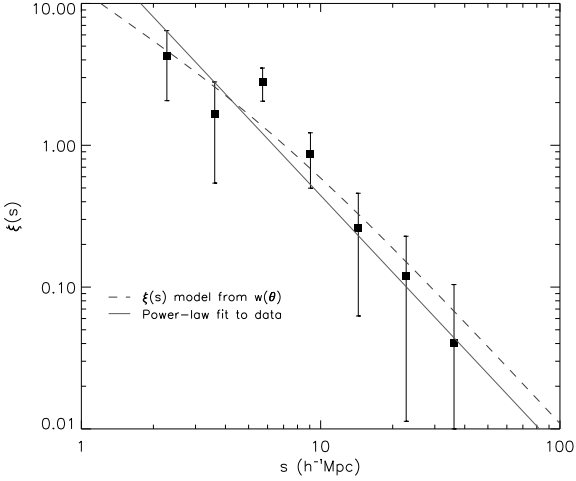}
 \caption{The redshift-space correlation function ($\xi(s)$) for the full sample of spectroscopically identified objects. The data-points show the correlation function of the spectroscopic sample of galaxies incorporating the S11, COSMOS, e04 and d05 fields. The solid line shows a best fit to the data points (with a fixed slope of $\gamma=1.8$), which is characterised by a correlation length of $s_o=6.4\pm0.8h^{-1}Mpc$. The dashed line is the $\xi(s)$ determined from the angular correlation function estimate of the real-space correlation function ($r_o=5.88h^{-1}Mpc$, $\gamma=1.83$ at $r>1h^{-1}Mpc$). This incorporates the coherent infall ($\beta=0.54$) and random pairwise velocity ($a=500\kps$) effects.}
 \label{xis}
\end{figure}

We use the correlation estimator given in equation~\ref{eq:wth} to determine $\xi(s)$, whilst errors are estimated using Poisson errors. The result is shown in figure~\ref{xis} (filled square points).  We fit the $\xi(s)$ measurement with a single power-law (noting that the break used in the double-power laws previously lies below the range of our $\xi(s)$ result) and find a best-fit (using a fixed slope of $\gamma=1.8$) given by a clustering length of $s_o=6.4\pm0.8 h^{-1}Mpc$ (solid line).

We also show the expected $\xi(s)$ determined from the power-law form of $\xi(r)$ given by our estimate of $w(\theta)$ (dashed line). To do this, we take the power-law form and apply both coherent infall and random pairwise velocity effects. The coherent infall imprint on the correlation function is characterised by the infall parameter, $\beta$, which is given by:

\begin{equation}
\beta = \frac{\Omega_m(z)^{0.55}}{b} 
\end{equation}

\noindent where $\Omega_m(z)$ is the mass density at the required redshift and $b$ is again the sample bias \citep{kaiser87}. Using the value of $b=1.46$ from table~\ref{r0par}, this gives a value of the infall parameter of $\beta=0.54$. For the random pairwise velocities we use a value of $a=500\kps$ as a reasonable estimate of the random motions based on the 2dFGRS data \citep{hawkins03}. The expected $\xi(s)$ was then calculated using the relations given in \citet{hawkins03}. Based on this calculation, we see from figure~\ref{xis} that we find good agreement between the $\xi(s)$ of our observed spectroscopic sample and that determined from the photometric data, within the associated errors.

\section{Integrated Sachs Wolfe effect}
\label{sec:isw}

\subsection{Overview}

As described in the introduction the ISW effect is characterised by the energy boost that CMB photons experience as they cross temporally evolving gravitational potential wells in an accelerating Universe. The effect is therefore a potential tool for placing constraints on the acceleration of the Universe as characterised by the cosmological constant, $\Lambda$. Given the effect's large scale, the production of all-sky CMB maps from the WMAP experiment has made it possible to attempt measures of the effect through the cross-correlation of galaxies (as tracers of the large gravitational potentials) with the CMB. Further to this, the large scale nature of the effect lends itself well to the use of photometrically selected galaxy populations as detailed redshift information is unnecessary. In this vein, several authors (e.g. \citealt{fosalba03, scranton03, cabre06, rassat07}) have used a number of galaxy samples to attempt measurements of the ISW signal in the WMAP 1st and 3rd year data. The samples used have mostly been photometrically selected LRGs at redshifts of $z<0.6$ and simple magnitude selected samples.

We now use our ELG sample to attempt to measure the ISW effect in WMAP 5-year data at our sample redshifts of $z\approx0.3$, $z\approx0.5$ and $z\approx0.7$. As stated much of the ISW work done thus far has been with LRGs and magnitude cut samples at $z<0.6$. Our use of the ELG samples provides the benefit of extending to greater redshifts, whilst also using an alternative galaxy population. This in itself has benefits and draw-backs. Firstly, the measured signal will be heavily dependant on the bias of the sample (i.e. how well the sample traces the dark matter structure and hence the gravitational potential). Given that the ELGs are less clustered than the LRG samples used thus far, we therefore expect to measure a weaker signal, making the measurement potentially more difficult. The potential gain in the low clustering strengths of the ELG samples however, is that they are less likely to reside in rich clusters and so we may expect the ISW signal to be less affected by the SZ effect produced as CMB photons pass through hot intra-cluster gas. This potentially provides an interesting alternative to the highly clustered LRG samples used in a number of the previous studies.

\subsection{Data}

For this cross-correlation, we have used the W, V, Q and K band temperature maps from the WMAP 5-year data release \citep{hinshaw}. We use the full-resolution maps in all cases. Before performing the cross-correlation we apply two masks to the data. The first is the WMAP KP0 mask \citep{bennett} which removes the majority of the galactic (Milky Way) foreground and is the most rigorous mask provided by the WMAP team. Secondly we mask the data to match the coverage of our SDSS DR6 galaxy samples, which is described further below.

Pixelised sky-density maps are constructed from each of the three galaxy samples using the HEALPIX software. These are constructed with a resolution identical to the WMAP temperature maps characterised by the HEALPIX parameter $NSIDE=512$ (pixel width$\approx7\arcmin$). We then limit our galaxy sample to incorporate only the contiguous north galactic pole region of the SDSS. Thus our sample is limited to $100^\circ < R.A. < 270^\circ$ and stripes 39, 42 and 43 are also excluded.

\subsection{Method}

Following the work of \citet{fosalba03} and \citet{cabre06}, we use the estimate the cross-correlation of the galaxy and WMAP data as the expectation value of the product of the galaxy over-density, $\delta_g=\frac{\rho_g-\overset{\_}{\rho}_g}{\overset{\_}{\rho}_g}$ and the normalised CMB anisotropy temperature, $\Delta T=T-\overset{\_}{T}$ as a function of the angular separation, $\theta$. This is given by:

\begin{equation}
w_{Tg}(\theta)=\frac{\sum_{i,j}\Delta T(\theta_i)\delta_g(\theta_j)}{n_{\Delta T}n_{\delta_g}}
\end{equation}

Again following \citet{fosalba03} and \citet{cabre06}, the form of the ISW as probed by a given galaxy population can be expressed by the following Legendre polynomial expansion:

\begin{equation}
w^{ISW}_{TG}(\theta) = \sum_{l} \frac{2l+1}{4\pi}p_l(\cos\theta)C^{ISW}_{GT}(l)
\end{equation} 

$C^{ISW}_{GT}(l)$ is simply the ISW/galaxy population power spectrum as given by:

\begin{equation}
C^{ISW}_{GT}(l) = \frac{4}{(2l+1)^2}\int W_{ISW}(z)W_G(z)\frac{H(z)}{c}P(k)dz
\end{equation}

\noindent where P(k) is the mass power spectrum and $W_{ISW}(z)$ and $W_{G}(z)$ are given by:

\begin{equation}
W_{ISW}(z) = 3\Omega_m\left(\frac{H_0}{c}\right)^2\frac{d[D(z)/a]}{dz}
\end{equation}

\begin{equation}
W_{G}(z) = b(z)\phi(z)D(z)
\end{equation}

\noindent where $D(z)$ is the linear growth rate and $b(z)$ is the bias of the galaxy population (taken from section~\ref{sec:clustering}). $\phi(z)$ is the galaxy selection function, set from the $n(z)$ distribution of each of the galaxy samples.

\subsection{Results \& Error Analysis}

\begin{figure}
 \centering
 \includegraphics[width=80.mm]{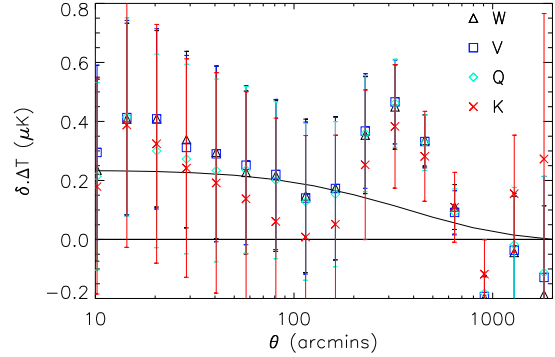}
 \caption[{\small Cross correlation between the low-redshift galaxy sample and the WMAP V band data.}]{\small Cross correlation between the low-redshift galaxy sample and the WMAP V band data. The solid line shows the predicted model. Errors are field-to-field based on splitting the data sample into 16 distinct segments.}
 \label{ISWlow}
\end{figure}

\begin{figure}
 \centering
 \includegraphics[width=80.mm]{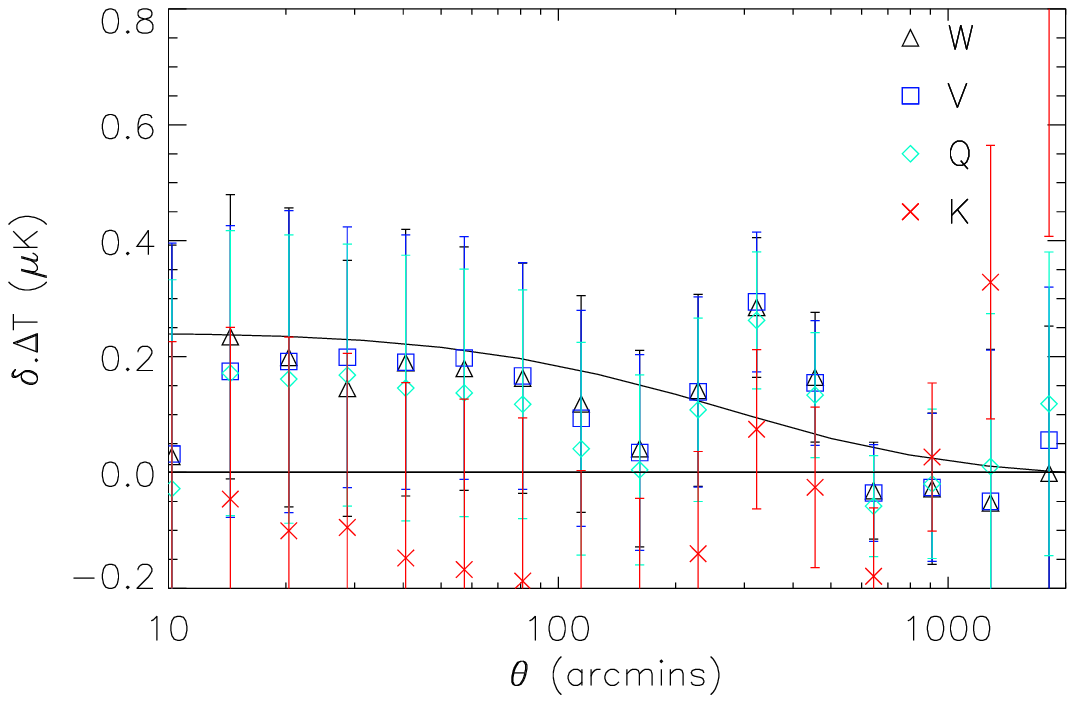}
 \caption[{\small Cross correlation between the mid-redshift galaxy sample and the WMAP V band data.}]{\small{As in figure~\ref{ISWlow} but with our mid-redshift sample of emission line galaxies.}}
 \label{ISWmid}
\end{figure}

\begin{figure}
 \centering
 \includegraphics[width=80.mm]{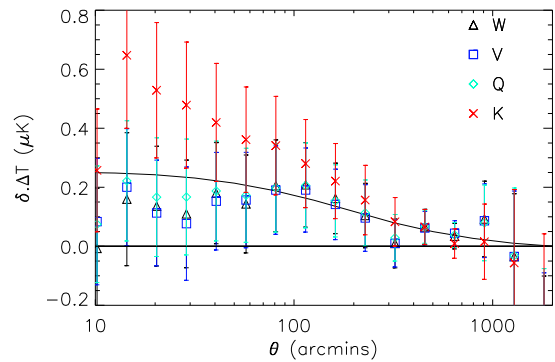}
 \caption[{\small Cross correlation between the high-redshift galaxy sample and the WMAP V band data.}]{\small{As in figure~\ref{ISWlow} but with our high-redshift sample of emission line galaxies.}}
 \label{ISWhigh}
\end{figure}

We perform the cross correlation using the {\small NPT} (N-point spatial statistic) software \citep{gray} with the weighting for each pixel given by the galaxy density, $\delta_g$ and the CMB anisotropy temperature, $\Delta T$. The results are shown in figures~\ref{ISWlow}~to~\ref{ISWhigh} for four WMAP bands: W, V, Q and K. We also plot the predicted result using predictions based on equation 6 of \citet{cabre06}.

We estimate the errors on the cross-correlation analysis using field-to-field errors. For this purpose, we split the studied region into 16 approximately equal area sub-fields and re-calculate the cross-correlation within each sub-field. The error in each angular bin is thus estimated as $\sigma/\sqrt{n}$, where $\sigma$ is the standard deviation across the sub-fields and $n$ is the number of sub-fields.

Summing over all bins at $\theta<100\arcmin$ we find amplitudes for $w_{Tg}(<100\arcmin)$ in the WMAP W-band of $0.25\pm0.27\mu K$, $0.17\pm0.20 \mu K$ and $0.17\pm0.16 \mu K$ for the low, mid and high redshift samples respectively. Similar results are obtained with the V and Q bands, whilst the K-band (which has a greater level of galactic contamination and a lower resolution) is less consistent, giving signals of $w_{Tg}(<100\arcmin)=0.13\pm0.36 \mu K$, $-0.16\pm0.29 \mu K$ and $0.38\pm0.18 \mu K$ for the low, mid and high redshift samples respectively.

\begin{figure}
 \centering
 \includegraphics[width=80.mm]{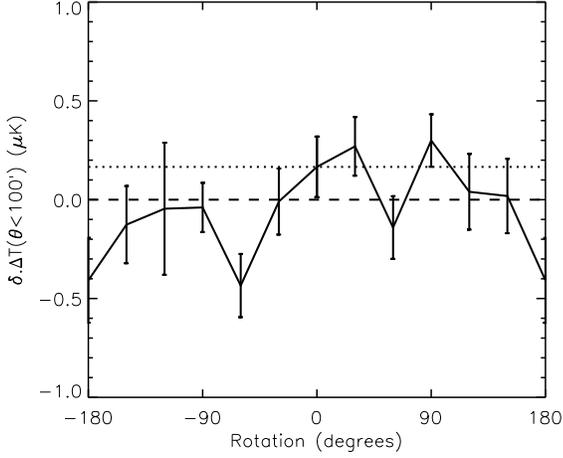}
 \caption[{\small Cross-correlation signal, $w_{Tg}(<100\arcmin)$, for the high-redshift galaxy sample as a function of rotation of the WMAP 5yr data in galactic longitude.}]{\small{Cross-correlation signal, $w_{Tg}(<100\arcmin)$, for the high redshift galaxy sample as a function of rotation of the WMAP 5yr data in galactic longitude. The plotted errors are field-to-field errors based on the segmentation of the data into 16 distinct regions and the dotted line shows the measurement from the non-rotated data.}}
 \label{ISWrot}
\end{figure}

We also evaluate the significance of the observed correlation by repeating the cross-correlation with rotated realisations of the WMAP data. This method uses the data itself in place of random realisations by rotating the masked WMAP data in $30^\circ$ steps in galactic longitude. We note at this point that rotating in R.A. would lead to the galactic plane entering the field of view and although the galactic plane region is masked, it would reduce the number pixels in the analysis significantly. For consistency, we also rotate the WMAP Kp0 mask before applying it to the galaxy density map. The result of this treatment, using the high redshift sample, is given in figure~\ref{ISWrot}. Here the $w_{Tg}(<100\arcmin)$ signal is plotted as a function of rotation of the WMAP data through a full $360^\circ$ in galactic longitude. The dotted line shows the non-rotated signal. Again we see that the positive signal that we see in the data does not appear statistically significant, with the rotated results showing a large amount of scatter around $w_{Tg}(<100\arcmin)=0\mu K$ and two of the results (at $30^\circ$ and $90^\circ$) showing more significant positive correlation than the non-rotated result.

\begin{figure}
 \centering
 \includegraphics[width=80.mm]{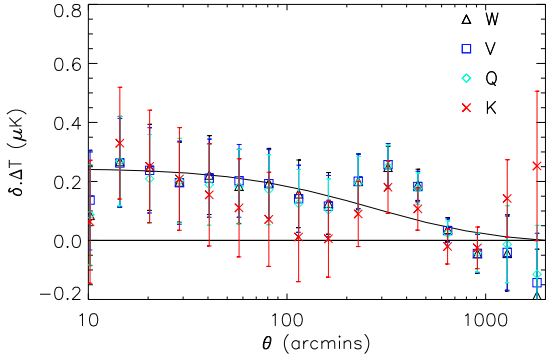}
 \caption[{\small Cross correlation result averaged across the $3\times16$ redshift/segment samples.}]{\small{Cross correlation result averaged across the $3\times16$ redshift/segment samples.}}
 \label{ISWall}
\end{figure}

\begin{figure}
 \centering
 \includegraphics[width=80.mm]{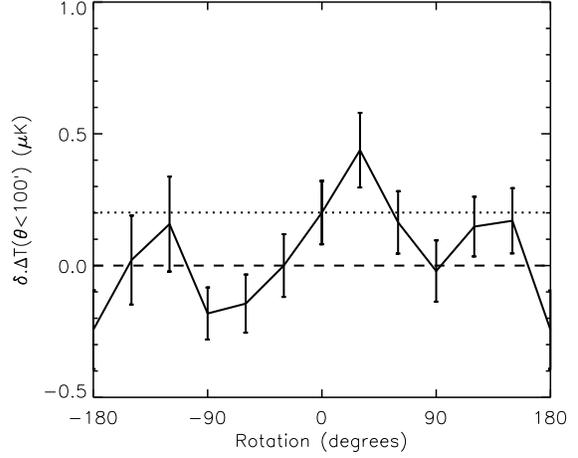}
 \caption[{\small Cross-correlation signal, $w_{Tg}(<100\arcmin)$, for the combined low, mid and high redshift galaxy sample as a function of rotation of the WMAP 5yr data in galactic longitude.}]{\small{Cross-correlation signal, $w_{Tg}(<100\arcmin)$, for the combined low-, mid- and high-redshift galaxy sample as a function of rotation of the WMAP 5yr data in galactic longitude. The plotted errors are field-to-field errors based on the segmentation of the data into $3\times16$ distinct regions. Again the result from the non-rotated data is shown by the dotted line.}}
 \label{allISWrot}
\end{figure}

We now attempt to improve our statistics by combining the low, mid and high redshift results. Figure~\ref{ISWall} shows the mean of the $3\times16$ separate cross-correlation results. The errors are again given by the field-to-field errors, this time across the whole 48 sample set. Again we see a positive signal that appears to show some agreement with the model. In the W, V, Q and K WMAP bands we derive signals of $w_{Tg}(<100\arcmin)=(0.20\pm0.12)\mu K$, $(0.20\pm0.12)\mu K$, $(0.18\pm0.12)\mu K$ and $(0.11\pm0.16)\mu K$ respectively. Repeating the rotation analysis (figure~\ref{allISWrot}), but with the combined sample, we again find significant scatter about $w_{Tg}(<100\arcmin)=0$. Indeed, a stronger signal is again found at some rotation angles than at the zero position.

Comparing this analysis to previous results, \citet{cabre06} obtained a signal at $100\arcmin$ of $w_{Tg}=0.6\pm0.3\mu K$ using a sample of $\overset{\_}{z}\approx0.5$ LRGs and a signal of $w_{Tg}=\approx0.65\pm0.2\mu K$ with a $20<r<21$ magnitude selected sample with a median redshift of $\overset{\_}{z}=0.28$. In addition, \citet{fosalba04} claimed a detection of $0.35\pm0.14\mu K$ at angular scales of $\theta=4\deg-10\deg$ using APM galaxies, but found their signal was dominated by the SZ component at scales of $\theta<4\deg$. Our observed signal is consistent with the weaker signal expected of the ELG samples, based on both the model predictions and comparison with correlations based on more strongly clustered populations. However, the estimated statistical and particularly the systematic errors mean that no detection of the ISW effect can be claimed at this point.

\section{Conclusions}
\label{sec:summary}
Colour selected samples are an extremely useful tool in modern astronomy and cosmology. They offer a cheap route to large galaxy redshift surveys, facilitating investigations in several key areas of interest, including the study of Dark Energy via both the Baryon Acoustic Oscillations and the Integrated Sachs Wolfe. With this in mind we have developed photometric selections with which to identify galaxies in three broad redshift ranges characterised by $\overset{\_}{z}=0.29\pm0.05$, $\overset{\_}{z}=0.44\pm0.08$ and $\overset{\_}{z}=0.65\pm0.21$. Applying these to SDSS data, we are able to select galaxy samples covering the given redshift ranges down to magnitudes of $i\approx20.5$.

The calibration of the high-redshift sample has been performed, using the AAOmega spectrograph at the AAT to provide spectroscopic redshifts for this sample. From these observations, we have shown that it is possible to target and spectroscopically identify $z>0.5$ star-forming galaxies with integration times of just 1 hour on the 4m AAT. The results of this have shown our high-redshift selection to work well, giving a close match to the expected redshift distribution, with $\overset{\_}{z}=0.66\pm0.23$. Further, this has proven that our selection can be a firm basis on which to conduct large scale spectroscopic surveys of $z>0.5$ emission line galaxies.

We have investigated the clustering properties of our three photometric samples using the angular correlation function. By fitting the angular correlation function using Limber's formula, we estimate the real-space clustering properties of the samples and find that both the low and high redshift populations are best fit by double power laws, whilst the mid-redshift population seems best fit by a single power-law. From these calculations, we estimate clustering lengths (at $r>0.5\hmpc$) of $2.65\pm0.12\hmpc$, $3.62\pm0.12\hmpc$ and $5.8\pm0.12\hmpc$ for the low, mid and high redshift samples respectively. Further to this, from our spectroscopic observations we measure a clustering length for the high redshift sample of $6.4\pm0.8\hmpc$ in agreement with the measurement based on the angular clustering measurement. Comparing these clustering measurements with comparable populations of late-type galaxies from 2dF data, we note that the high redshift sample appears to have an unexpectedly high clustering strength.

The development of photometric redshift selections has a number of scientific applications, not least the evaluation of the ISW effect. We have used our photometric selection to evaluate the ISW effect in the region of the SDSS by cross correlating the density fluctuations in our galaxy distributions with the CMB anisotropies from the WMAP 5yr data. The results obtained using all three datasets show a positive correlation in accordance with the predicted model. However, none of these three prove to be significant, with signals in the WMAP W-band of $(0.25\pm0.27)\mu K$, $(0.17\pm0.20)\mu K$ and $(0.17\pm0.16)\mu K$ for the low, mid and high redshift samples respectively. We attempt to improve the statistics by combining the three redshift samples, which results in a signal of $(0.20\pm0.12)\mu K$ when cross-correlated with the WMAP W-band, still only marginal ($1.67\sigma$). Also in tests of contamination by Êsystematics, we found similar results at arbitrary angles of rotation between the CMB data and the ELG samples which means that no detection of the ISW can be claimed above the random
and systematic noise in this analysis.

\section*{Acknowledgements}

We thank C. Wolf for supplying the COMBO-17 photometric redshift catalogue data and the staff of the Anglo-Australian Observatory for their work in operating the AAOmega facility during our observations. Specific thanks also go to Rob Sharp for assistance during the AAT observations. This paper has used data from both the SDSS and WMAP projects. Funding for the SDSS and SDSS-II has been provided by the Alfred P. Sloan Foundation, the Participating Institutions, the National Science Foundation, the U.S. Department of Energy, the National Aeronautics and Space Administration, the Japanese Monbukagakusho, the Max Planck Society, and the Higher Education Funding Council for England. The SDSS Web Site is http://www.sdss.org/. WMAP is the result of a partnership between Princeton University and NASA's Goddard Space Flight Center. Scientific guidance is provided by the WMAP Science Team. RMB and NPR acknowledge the support of a STFC PhD Studentships. SMC acknowledges the support of an Australian Research Council QEII Fellowship and a J G Russell Award from the Australian Academy of Science.

\label{lastpage}

\end{document}